\documentclass[
  aps,%
  12pt,%
  final,%
  notitlepage,%
  oneside,%
  onecolumn,%
  nobibnotes,%
  nofootinbib,
  superscriptaddress,%
  noshowpacs,%
  centertags]%
	      {revtex4}

\begin{document}

\selectlanguage{english}

\title{The Mass Function of Dark Halos in Superclusters and Voids}

\author{\firstname{E.~P.}~\surname{Kurbatov}}
\email{kurbatov@inasan.ru}
\affiliation{Institute of Astronomy of the Russian Academy of Sciences}

\begin{abstract}

A modification of the Press--Schechter theory allowing for presence of a
background large-scale structure (LSS) -- a supercluster or a void is
proposed. The LSS is accounted as the statistical constraints in form of linear
functionals of the random overdensity field. The deviation of the background
density within the LSS is interpreted in a pseudo-cosmological sense. Using the
constraints formalism may help us to probe non-trivial spatial statistics of
haloes, e.g. edge and shape effects on boundaries of the superclusters and
voids. Parameters of the constraints are connected to features of the LSS: its
mean overdensity, a spatial scale and a shape, and spatial momenta of higher
orders. It is shown that presence of a non-virialized LSS can lead to an
observable deviation of the mass function. This effect is exploited to build a
procedure to recover parameters of the background perturbation from the
observationally estimated mass function.

\bigskip\noindent
PACS numbers:\, 
\end{abstract}
\maketitle

\section{Introduction}

Mass function of \citet{Press--1974ApJ...187..425P} (hereafter PS) for dark
haloes depends on global cosmological parameters so does not incorporate the
presence of the background large-scale disturbances which can form
superclusters and voids. Presence of the LSS alter the matter density and can
change the growth rate of cosmological fluctuations. This problem was first
considered by \citet{Bond--1991ApJ...379..440B} and
\citet{Bower--1991MNRAS.248..332B} as the `two barrier' problem and improved by
\citet{Mo--1996MNRAS.282..347M}, \citet{Sheth--1999MNRAS.308..119S}, and
others. These models provided a connection between the mass function and the
background overdensity making possible to solve the inverse problem of
recovering the last, as was done by \citet{Munoz--2008MNRAS.391.1341M} and
\citet{Sheth--2011MNRAS.417.2938S}. In these models, however, the background
perturbation was considered spherical. Also there was not provided a
possibility to examine effects of transition between structures of different
densities, e.g. between superclusters and voids.

In this paper a modification of the PS theory is proposed allowing for a
background perturbation (name it `host') which is associated with the
supercluster or the void. The host perturbation is defined by a set of
statistical constraints for the cosmological perturbation field. Growth rate of
the random perturbations depends on the host overdensity and determines
dynamical age of the population of haloes. As a demonstration of applicability
of the model the inverse problem is considered in a simple case for recovering
the host overdensity.

Below, in 2nd Section described the modification of the PS theory. In the 3rd
Section considered an application of the theory to superclusters and
voids. Benefits and issues are discussed in the 4th Section.

\section{Press--Schechter theory with statistical constraints}

\subsection{Outline}
\label{sec:outline}

The key element of the excursion set theory is the assumption that the
virialized haloes form in a hierarchical stochastic process of absorbtion of
haloes of lower masses. Herewith, only those haloes are counted for mass
function which are on a top level of a hierarchical tree, i.e. are not
sub-haloes of any other halo. The stochastic nature of the process is governed
by properties of the initial cosmological fluctuations. Conditions for
fluctuations to form a virialized object may be implied to linear stage of
their evolution, and with a good approximation they are conditions for the
filtered overdensity only. A halo condensed when its characteristical linearly
evolved overdensity exceeded some threshold value given from the spherical
collapse model. Defining the statistical properties of the fluctuations, their
growth rate, and the threshold overdensity allows for completely determining of
the function of mass of the haloes for any redshift
\citep{Bond--1991ApJ...379..440B,Lacey--1993MNRAS.262..627L}.

The idea is necessary for this research to be done is to describe a large-scale
perturbation using formalism of statistical constraints in such a way so that
would be possible to define various parameters of the host, e.g. volume
averaged density, spatial scale, momenta of inertia etc. The cosmological
fluctuations, from which the sub-structures developed, should evolve on top of
the background modified by the host. Let's define $\phi$ as the residual
between a total overdensity field $\delta$ and overdensity of the host $\Delta$
(which is the ensemble mean for $\delta$, and will be defined in
Subsec. \ref{sec:constrained-correlation-function}):
\begin{equation}
  \phi(\vecb{r}) \equiv \delta(\vecb{r}) - \Delta(\vecb{r})  \;,
\end{equation}
where $\vecb{r}$ is the spatial position. Also define the field which is
filtered over a volume with characteristic scale $R$:
\begin{equation}
  \phi(\vecb{r}, R) \equiv
  \int \mathrm{d}^3r'\,W(|\vecb{r} - \vecb{r}'|, R)\,\phi(\vecb{r}')  \;,
\end{equation}
where $W(r, R)$ is the filtering function.
In a simple approximation all the interesting information on the features of
the host contains in variance of the filtered field
$\Sigma(\vecb{r}, R) \equiv \langle \phi^2(\vecb{r}, R) \rangle$. According to
excursion set theory, the hierarchical halo formation process can be
interpreted as a random process for the overdensity filtered over the scale $R$
acting as a parameter of the random process. The process starts at the
parameter $R = \infty$ moving toward $R = 0$. The mass function can be
expressed then via the distribution function for the parameter values when the
process first crossed a certain threshold. In presence of the constraints the
statistical homogeneity and isotropy of the fluctuations field may be broken
and spectral modes of the fluctuations may have non-trivial
correlations. Because of this, the excursion set approach in its standard form
may not work. This issue will be touched in Subsection \ref{sec:mass-df}.

In theories of PS class the evolution of amplitude of the fluctuations can be
accounted either as a time-dependent overdensity threshold or as evolution of
the variance. We will stick to the second scheme since it makes 
possible, in principle, to consider the non-linear corrections to the field
evolution. Exact calculation for evolution of the fluctuations' amplitude is
difficult even in lowest orders of the perturbation series approach.
For Gaussian field these calculations were
completed just up to 1-loop corrections, i.e. to 2-nd order of accuracy in
power spectrum \citep{Jain--1994ApJ...431..495J}, or up to 3-rd order in
expressions for filtered statistical momenta
\citep{Scoccimarro--1998MNRAS.299.1097S}. In a non-gaussian case the task
became more complicated since the field has not zero mean so the same precision
order requires at least twice as many integrals to get
\citep{Crocce--2006PhRvD..73f3519C,Crocce--2006PhRvD..73f3520C}. In our theory
we will restrict ourselves to a purely linear evolution, so the overdensity
will be proportional to the linear growth factor $D$:
\begin{equation}
  \label{eq:linear-evolution}
  \phi(z, \vecb{r}) = D(z, \vecb{r})\,\phi_\mathrm{L}(\vecb{r})  \;.
\end{equation}
Hereafter the `$\mathrm{L}$' index means values linearly evolved to the present
time with unity growth factor.

The linear evolution of perturbations settled on a large scale host overdensity
of both signs (the supercluster or void) can be represented in a
perturbation theory for the certain cosmology, which parameters are determined
by density of the host. In the overdense host the perturbations will attain
higher values of the growth factor than in global cosmology, in the underdense
host the last will be lower. Thus, the host overdensity is in charge of the age
of haloes population.

It is important to note that all spatial distributions below are defined in a
lagrangian coordinates comoving with the matter in modified background. We will
distinguish them from the coordinates comoving with the Hubble flow which
called the eulerian coordinates.

In this work the $\Lambda$CDM model is used with following parameters:
$\Omega_{\Lambda,0} = 0.7$, $\Omega_\mathrm{m,0} = 0.3$, $\sigma_8 = 0.9$. The
power spectrum is $P(k) \propto k T^2(k)$, where the transfer function is
computed using the {\sc cmbfast} code of
\citet{Seljak--1996ApJ...469..437S}. The units adopted are $h^{-1} \Msol$ for
mass, $h^{-1}$Mpc for length and $h^{-1} H_{100}^{-1}$ for time, where
$H_{100} = 100$~km~s$^{-1}$~Mpc$^{-1}$.

\subsection{Constrained correlation function for modes}
\label{sec:constrained-correlation-function}

To calculate the statistical characteristics such as the variance and the
spatial correlation function, we need the correlation function for modes of the
constrained field. It's convenient to make calculations using spherical modes
decomposition. To obtain the correlation function for amplitudes of the
spherical modes we will use approach of \citet{Hoffman--1992ApJ...384..448H},
which was proposed by them for plane waves.

The field $\delta(\vecb{r}) = \delta(r, \vecb{\omega})$%
\footnote{The index `$\mathrm{L}$' is dropped in this Subsection.}
can be decomposed to
spherical modes via transformation
\begin{equation}
  \delta(\vecb{r}) =
  \sqrt{\frac{2}{\pi}} \sum_{l=0}^\infty \sum_{m=-l}^l
  \int_0^\infty \mathrm{d}k\,
  k^2 j_l(kr)\,Y_{lm}(\vecb{\omega})\,\tilde{\delta}_{lm}(k)  \;,
\end{equation}
where $j_l$ is the radial Bessel's functions; $Y_{lm}$ is the spherical
functions with normalization
$\oint_{4\pi} \mathrm{d}\omega\,Y_{lm}^\ast Y_{l'm'} =
4\pi \delta_{ll'} \delta_{mm'}$.
Amplitude of spherical decomposition of the field, or image, is
\begin{equation}
  \tilde{\delta}_{lm}(k) =
  \sqrt{\frac{2}{\pi}} \int_0^\infty \mathrm{d}r\,r^2 j_l(kr)
  \oint_{4\pi} \mathrm{d}\omega\,Y_{lm}^\ast(\vecb{\omega})\,
  \delta(r, \vecb{\omega})  \;.
\end{equation}
The image of radially symmetric field $H(r)$ consists of only isotropic modes:
\begin{equation}
  \tilde{H}_{lm}(k) = \delta_{l0} \delta_{m0} \tilde{H}(k)  \;,
\end{equation}
where $\delta_{ij}$ is the Kronecker's delta, while
\begin{equation}
  \label{eq:image-symmetric}
  \tilde{H}(k)
  = \sqrt{\frac{2}{\pi}} \int_0^\infty \mathrm{d}r\,r^2 j_0(kr) H(r)  \;,
\end{equation}
\begin{equation}
  H(r)
  = \sqrt{\frac{2}{\pi}} \int_0^\infty \mathrm{d}k\,k^2 j_0(kr) \tilde{H}(k)
  \;.
\end{equation}

Let us write constraints in form of linear functionals, assuming kernels are
spherically symmetric and place their centers at origin. Using this conditions,
the functional for $\alpha$-th constraint can be written as
\begin{equation}
  \label{eq:constraining-functional}
  C_\alpha[\delta] = \int \mathrm{d}^3r\,H_\alpha(r)\,\delta(\vecb{r}) =
  4\pi \int_0^\infty \mathrm{d}k\,k^2 \tilde{H}_\alpha(k)\,
  \tilde{\delta}_{00}(k)  \;.
\end{equation}
The constraints itself are fixed by assigning certain values $C_\alpha$ to
these functionals. Also the kernels can depend on a set of parameters
characterizing the host, like a spatisl scale, derivations, or momenta.

Following \citet{Hoffman--1991ApJ...380L...5H,Hoffman--1992ApJ...384..448H} it
can be showed that the ensemble mean of the constrained field is
\begin{align}
  \label{eq:constrained-overdensity}
  &\Delta(\vecb{r})
  = \langle \delta(\vecb{r}) | \{C_\alpha\} \rangle  \;,  \\
  \label{eq:constrained-overdensity-image}
  &\tilde{\Delta}_{lm}(k)
  = \langle \tilde{\delta}_{lm}(k) | \{C_\alpha\} \rangle
  = \delta_{l0} \delta_{m0}
  Q_{\alpha\beta}^{-1}\,C_\alpha\,\tilde{\xi}_\beta(k)  \;,
\end{align}
where $C_\alpha$ are the values of the constraining functionals;
$\tilde{\xi}_\alpha(k)$ are images of cross-correlation function between the
field and $\alpha$-th constraint:
\begin{equation}
  \label{eq:field-constr-correlator}
  \xi_\alpha(\vecb{r}) = \langle \delta\,C_\alpha[\delta] \rangle
  \;,\qquad
  \tilde{\xi}_\alpha(k) = \tilde{H}_\alpha(k)\,P(k)  \;;
\end{equation}
and $Q_{\alpha\beta}^{-1}$ is the inversion of the constraints' correlation
matrix
\begin{equation}
  \label{eq:constr-constr-correlator}
  Q_{\alpha\beta} = \langle C_\alpha[\delta]\,C_\beta[\delta] \rangle
  = 4\pi \int_0^\infty \mathrm{d}k\,k^2
  \tilde{H}_\alpha(k)\,\tilde{H}_\beta(k)\,P(k)  \;;
\end{equation}
we also can write $Q_{\alpha\beta} = C_\alpha[\xi_\beta]$. It can be showed
that the pair correlation function for the constrained ensemble of modes is
determined by the residuals:
\begin{equation}
  K_{lml'm'}(k, k')
  = \left\langle \tilde{\phi}_{lm}(k)\,\tilde{\phi}_{l'm'}(k') \right\rangle
  = \left\langle
  \left(\tilde{\delta}_{lm}(k) - \tilde{\Delta}_{lm}(k)\right)
  \left(\tilde{\delta}_{l'm'}(k') - \tilde{\Delta}_{l'm'}(k')\right)
  \right\rangle  \;,
\end{equation}
where averaging performed over the non-constrained ensemble. The
non-constrained field is delta-correlated, which means
\begin{equation}
  \label{eq:non-constr-correlation}
  \left\langle \tilde{\delta}_{lm}(k)\,\tilde{\delta}_{l'm'}(k') \right\rangle
  = \delta_{ll'} \delta_{mm'}
  \frac{\delta_\mathrm{1D}(k - k')}{4\pi k k'}\,P(k)  \;,
\end{equation}
where $\delta_\mathrm{1D}$ is the one-dimensional Dirac's
delta-function. Substituting the definition for ensemble mean field we have
\begin{equation}
  \label{eq:correlation-function}
  K_{lml'm'}(k, k') =
  \delta_{ll'} \delta_{mm'}\,
  \frac{\delta_\mathrm{1D}(k - k')}{4\pi k k'}\,P(k)
  - \delta_{l0} \delta_{m0} \delta_{l'0} \delta_{m'0}\,
  Q_{\alpha\beta}^{-1}\,\tilde{\xi}_\alpha(k)\,\tilde{\xi}_\beta(k')  \;.
\end{equation}
As can be seen, the presence of constraints leads to fact that the modes are
correlated. This changes the spatial statistics of the modes, particularly
induces the spatial dependency of the variance. General expression for variance
of the filtered field is
\begin{equation}
  \Sigma(\vecb{r}, R) = \langle \phi^2(\vecb{r}, R) \rangle  \;,
\end{equation}
where the filtering in spherical decomposition appears in the form
\begin{equation}
  \label{eq:filtered-field}
  \phi(\vecb{r}, R) = 4\pi \int_0^\infty \mathrm{d}k\,k^2 j_0(kr)\,
  \tilde{W}(k, R)\,\tilde{\phi}_{00}(k)  \;.
\end{equation}
The filter chosen is 'top-hat':
\begin{equation}
  \label{eq:top-hat-real}
  W(r, R) = \frac{3}{4\pi R^3}\,\theta(1 - r/R)  \;,
\end{equation}
\begin{equation}
  \label{eq:top-hat-image}
  \tilde{W}(k, R)
  = \frac{3}{(2\pi)^{3/2}}\,\frac{\sin kR - kR \cos kR}{(kR)^3}  \;.
\end{equation}
Substituting (\ref{eq:correlation-function}) into the general expression we
obtain
\begin{equation}
  \label{eq:variance}
  \Sigma(\vecb{r}, R)
  = S(R)
  - Q_{\alpha\beta}^{-1}\,\xi_\alpha(\vecb{r}, R)\,\xi_\beta(\vecb{r}, R)  \;,
\end{equation}
where $S(R)$ denoted the variance of the non-constrained field,
\begin{equation}
  \label{eq:variance-non-constrained}
  S(R) = 4\pi \int_0^\infty \mathrm{d}k\,k^2 \tilde{W}^2(k, R)\,P(k)  \;,
\end{equation}
and $\xi_\alpha(\vecb{r}, R)$ is the filtered cross-correlator
(\ref{eq:field-constr-correlator}) defined in the same way as in
Eq. (\ref{eq:filtered-field}). Because of the constraints the variance becomes
dependent on the spatial position. The magnitude of corrections in
Eq. (\ref{eq:variance}) decays to zero when $r \to \infty$. We also may expect
this behavior when $R \to \infty$.

\subsection{Constraining kernels and host profile}
\label{sec:kernels}

The amplitude of the mean constrained field or mean overdensity profile
is the linear combination of the cross-correlators between the field and the
constraints. Choosing appropriate constraining kernel it is possible to set a
particular profile for the host. Using set of the kernels, the mean field can
be represented as a combination of some basis profiles. Again, the constraints
can be thought of in terms of spatial momenta of the host's density
distribution, i.e. average value, moment of inertia, etc.

At the first time consider the single constraint with the top-hat kernel,
$H(r, R_\mathrm{H}) \equiv W(r, R_\mathrm{H})$. It is easy to see that value of
the constraining functional (\ref{eq:constraining-functional}) with this
kernel fixates the overdensity value averaged over a sphere of radius
$R_\mathrm{H}$. Several overdensity constraints applied to different
scales $R_{\mathrm{H},\alpha}$ may be combined to obtain more complex
profile. In this case the constraining kernels set should be defined as
\begin{equation}
  \label{eq:overdensity-constraints}
  H_\alpha(r) \equiv H(r, R_{\mathrm{H},\alpha})  \;.
\end{equation}

The kernels may be choosed to fixate the spatial momenta of different orders:
\begin{equation}
  \label{eq:momenta-constraints}
  H_\alpha(r) \equiv r^\alpha H(r, R_\mathrm{H})  \;.
\end{equation}

Yet another possible way to use constraints is to fixate values of the spatial
derivations of the host:
\begin{equation}
  \begin{aligned}
  C_\alpha[f] \equiv C\left[\Dfrac{^\alpha f}{r^\alpha}\right]
  = &\int \mathrm{d}^3r\,H(r, R_\mathrm{H})\,\Dfrac{^\alpha f(r)}{r^\alpha}  \\
  = &(-1)^\alpha \int \mathrm{d}^3r\,
  \DPfrac{^\alpha H(r, R_\mathrm{H})}{r^\alpha}\,f(\vecb{r})  \;.
  \end{aligned}
\end{equation}
The set of the kernels in this case is defined as
\begin{equation}
  \label{eq:derivations-constraints}
  H_\alpha(r) \equiv (-1)^\alpha\,\DPfrac{^\alpha H(r, R_\mathrm{H})}{r^\alpha}
  \;.
\end{equation}
Using reccurence relations for radial Bessel functions $j_l$ and their
derivations, it can be shown that all the differential constraints of the odd
order turn to zero because of symmetry. Images for the second and fourth
derivations of the kernel are expressed as
\begin{equation}
  \tilde{H}_2(k) = - \frac{k^2}{3}\,\tilde{H}(k)  \;,\quad
  \tilde{H}_4(k) = \frac{k^4}{5}\,\tilde{H}(k)  \;.
\end{equation}
In this approach the constraints actually fix the moments in the momentum space
of the convolved field.

All the examples above were based on the spatially symmertic kernel
$H(r, R_\mathrm{H})$. Hence, the ensemble average profile $\Delta(r)$,
Eq. (\ref{eq:constrained-overdensity}), has the same property.

\begin{figure*}
  \begin{center}
    \includegraphics{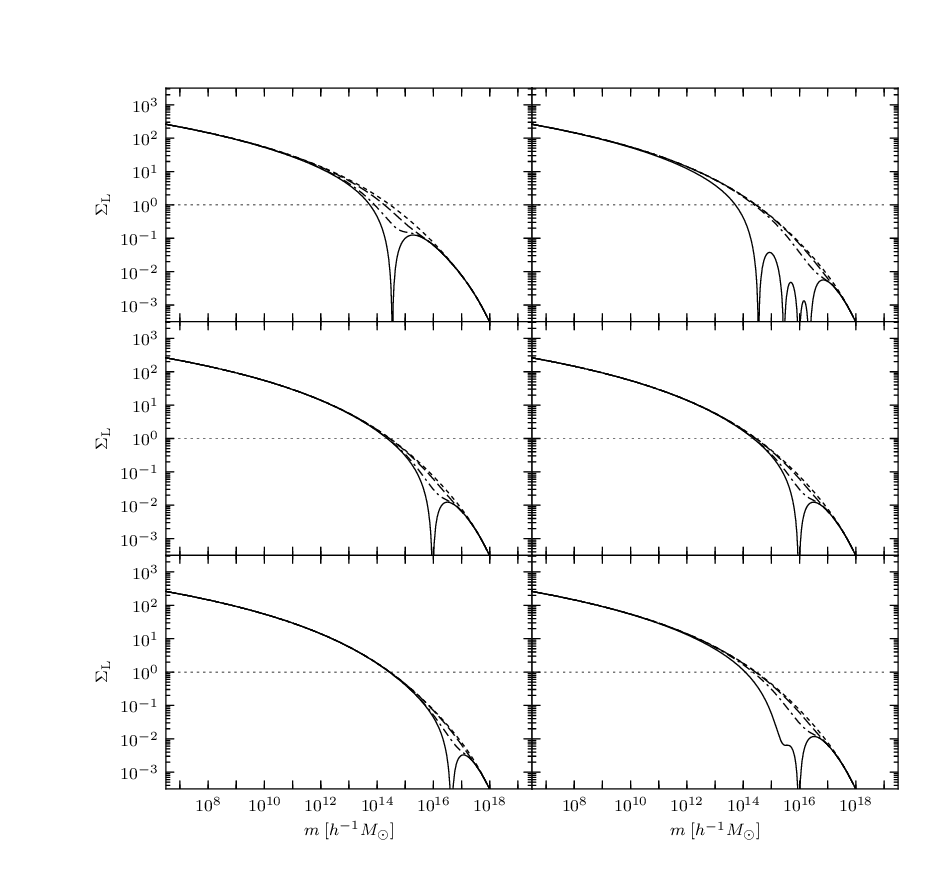}
  \end{center}
  \caption{Variance as a function of mass for different sets of
    constraints. {\em Left column.} The single top-hat constraint is used. From
    top to bottom: $R = 10$, $R = 30$, and $R = 50 h^{-1}$Mpc correspondingly.
    {\em Right column.} Top figure: used four top-hat constraints having scales
    $R = 10$, $20$, $30$, $40 h^{-1}$Mpc. Middle figure: used derivation
    constraint of the kind (\ref{eq:derivations-constraints}) for
    $R = 30 h^{-1}$Mpc. Bottom figure: used momenta constraints of the kind
    (\ref{eq:momenta-constraints}) for $R = 30 h^{-1}$Mpc.
    At all plots solid line is for $r = 0$, long-dash-dotted line is for
    $r = R/2$, long-dashed line is for $r = R$, and short-dashed represents
    variance for non-constrained field. In the case of multiple top-hat
    constraints (right top figure) $R = 40 h^{-1}$Mpc. Thin short-dashed line
    stands for unity level, the nominal level of non-linearity.}
  \label{fig:variance}
\end{figure*}

From the Eq. (\ref{eq:variance}) it follows that the constraining kernels (but
not the values of the constraining functionals) are responsible for
corrections to the variance, which is represented on
Fig. \ref{fig:variance}. On each plot the curves correspond to a fixed radial
position inside the host. The farther from the center of the host, the lesser
its influence on the filtered field, then the variance turns to its
non-constrained form. In the case when filtering spheres are adjusted to the
center of the host ($r = 0$), the variance falls to zero for a certain
filtering scale. This is the result of using the top-hat constraining kernel,
when a value of the filtered overdensity is fixed, i.e. the filtered field
loses randomness exactly on the scale $R = R_\mathrm{H}$.
The several dips appeared when several constraints of a kind
(\ref{eq:overdensity-constraints}) are implemented.
It was found that applying the derivation constraints
(\ref{eq:derivations-constraints}) does not cause differences from the case of
the single overdensity constraint (\ref{eq:overdensity-constraints}), even for
higher orders of derivating (not shown on the Fig. \ref{fig:variance}).

We can go the opposite way, to define the basis profiles then obtain the
corresponding set of constraints. Namely, let the cross-correlators
$\xi_\alpha(r)$ be the basis profiles. Given values $C_\beta$ of the
constraining functionals, we can express
\begin{equation}
  \label{eq:constrained-overdensity-real}
  \Delta(r) = Q_{\alpha\beta}^{-1} C_\beta \xi_\alpha(r)  \;,
\end{equation}
where $Q_{\alpha\beta} = C_\alpha[\xi_\beta]$. For example, the profile
$\Delta(r)$ could appear from a numerical computations as given, then a single
constraint could be generated via this scheme.

\begin{figure}
  \begin{center}
    \includegraphics{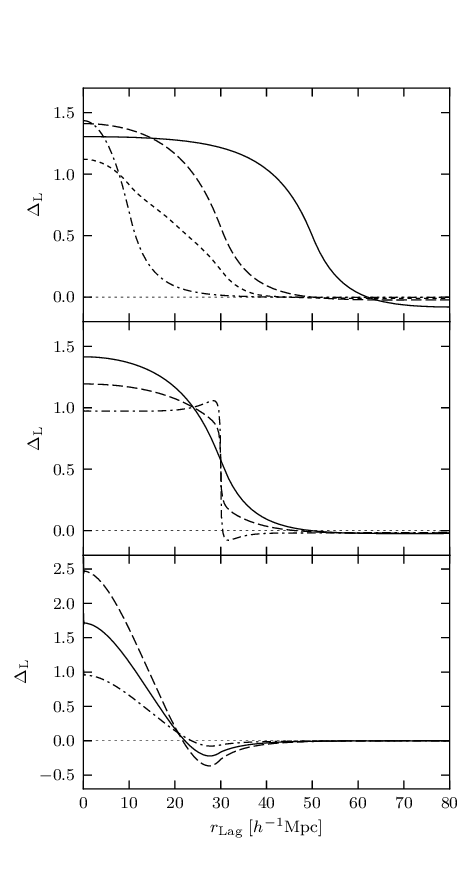}
  \end{center}
  \caption{Mean overdensity as a function of lagrangian radial
    coordinate. These profiles are obtained for different sets of constraints
    listed in Subsection \ref{sec:kernels}.
    {\em Top panel.} A single top-hat constraint for three values of the scale
    radius $R_\mathrm{H}$: $50$ (solid line), $30$ (long-dashed line), and
    $10 h^{-1}$Mpc (dot-dashed line). Used the same value of the averaged
    overdensity $1$. With a short-dashed line depicted a distribution with a
    combination of four top-hat constraints having different scale radii:
    $10$, $20$, $30$, $40 h^{-1}$Mpc and corresponding averaged
    overdensities $1$, $0.75$, $0.5$, and $0.25$.
    {\em Middle panel.} Derivations constraints of the kind
    (\ref{eq:derivations-constraints}) for $\alpha = 0, 2$ and 
    $R_\mathrm{H} = 30 h^{-1}$Mpc, where $C_0 = 1$, and $C_2$ values are $0$
    (solid line), $-0.25$ (long-dashed line), $-0.5$ (dot-dashed line).
    {\em Bottom panel.} Momenta constraints of the kind
    (\ref{eq:momenta-constraints}) for $\alpha = 0, 1$ and
    $R_\mathrm{H} = 30 h^{-1}$Mpc, where $C_0 = 0.1$, and $C_1$ values are $0$
    (solid line), $-1$ (long-dashed line), $1$ (dot-dashed line).}
  \label{fig:host-overdensity}
\end{figure}

Profiles resulting from constraints given above are shown on
Fig. \ref{fig:host-overdensity}. In all the cases, except when the momenta
constraints used, the volume averaged overdensity (denote it $C_0$) was set to
unity. On top panel the profile resulting from applying four overdensity
constraints (\ref{eq:overdensity-constraints}) is presented demonstrating the
possibility to model structures having complex form. The computed profiles
exhibited moderate variability between nearly zero and unity levels. In
contrast, when the momenta constraints were applied, the amplitude of variation
was much higher though $C_0$ was set to $0.1$. This suggests that the momenta
constraints (\ref{eq:momenta-constraints}) are hardly suitable for the
supercluster profile fitting as the overdensity profile they gave has too
strong features to fit.

It is important to note that all the spatial distributions mentioned above are
actually defined as a functions of lagrangian coordinates, i.e. comoving with
the matter of the growing host. The mass conservation during this process
causes displacements of lagrangian elements. Mapping of the lagrangian
coordinates to the eulerian ones, i.e. comoving with the Hubble flow, could be
followed tracking the evolution of the `fluid' element in a perturbed
gravitational field \citep{Hui--1996ApJ...471....1H}. When the perturbation,
the host in our case, has spherical symmetry, the mapping could be established
trivially:
\begin{equation}
  \label{eq:lagrangian-to-eulerian}
  r_\mathrm{eul}^3(z) = 3 \int_0^{r_\mathrm{lag}} \mathrm{d}r\,
  \frac{r^2}{1 + \Delta(z, r)}  \;.
\end{equation}

\subsection{Linear growth of perturbations}
\label{sec:growth-factor}

The evolution of perturbations inside an overdense or underdense region can be
examined in a pseudo-cosmological notation, if parameters of such notation are
chosen appropriately
\citep{Silk--1977A&A....59...53S,Peebles--1993ppc..book.....P}.
The parameters are: scale factor, Hubble constant, critical density, and
density parameters. For analysis of their dependency on the perturbation
overdensity see \citet{Martino--2009MNRAS.394.2109M}. The brief description is
following. The present day value of the scale factor of perturbation $a'_0$ is
determined by the present day amplitude of the region $\overline{\Delta}_0$ as
$a'_0 = (1 + \overline{\Delta}_0)^{-1/3}$ while $a' \approx a$, the global
scale factor, at early epoch. The Hubble constant $H'_0$ can be determined
implicitly, using these boundary conditions with equation
\begin{equation}
  \frac{a}{a'}\,\Dfrac{a'}{a}
  = \frac{H'(a'/a'_0, \Omega'_\mathrm{m,0}, \Omega'_{\Lambda,0})}%
  {H(a, \Omega_\mathrm{m,0}, \Omega_{\Lambda,0})}  \;,
\end{equation}
where the present day density parameters of perturbation are
\begin{equation}
  \label{eq:pseudo-cosmological-density-parameters}
  \Omega'_\mathrm{m,0} = (1 + \overline{\Delta}_0)\,
  \frac{\Omega_\mathrm{m,0}}{(H'_0/H_0)^2}
  \quad\text{and}\quad
  \Omega'_{\Lambda,0} = \frac{\Omega_{\Lambda,0}}{(H'_0/H_0)^2}  \;.
\end{equation}
The values found have to be substituted to the linearized equation for
overdensity, then the last will be defined relatively to the perturbed
background \citep{Martino--2009MNRAS.394.2109M}. The linear growth factor $D$
inside the spherical perturbation can be determined via equation
\begin{equation}
  a'H'\Dfrac{}{a'} \left( a'H'\Dfrac{D}{a'} \right) +
  2 a'H'\Dfrac{D}{a'} -
  \frac{3 \Omega'_\mathrm{m,0} H'^2_0}{2 (a'/a'_0)^3}\,D = 0  \;.
\end{equation}
The initial condition is the growing mode condition $D \approx a'$ at early
epoch, independing on $\overline{\Delta}_0$.

As we are building the local theory, the value of $\overline{\Delta}_0$ should
be interpret as a present day non-linear amplitude of the host profile averaged
over the volume of the small scale perturbation. However, the statistical
constraints applied to the cosmological perturbations result in the host
profile at high redshift. Instead of following non-linear evolution of the
host, assume the linear approximation (\ref{eq:linear-evolution}). The actual
host profile at a given redshift will be then
\begin{equation}
  \Delta(z, r) = D(z; \overline{\Delta}_0 = 0)\,\Delta_\mathrm{L}(r)  \;,
\end{equation}
where $D(z; \overline{\Delta}_0 = 0)$ is the linear growth factor for the
global cosmology, and $\Delta_\mathrm{L}(r)$ is the mean overdensity profile
(\ref{eq:constrained-overdensity}). Further, to obtain $\overline{\Delta}_0$
the averaging procedure should be applied to this profile at redshift zero.

We assume that the variance of the filtered perturbations (\ref{eq:variance})
seeded inside the host is also governed by the local linear law
\begin{equation}
  \Sigma(z, r, R) = D^2(z)\,\Sigma_\mathrm{L}(r, R)  \;,
\end{equation}
where $D$ is the local growth factor inside the host computed for the
corresponding amplitude $\overline{\Delta}_0$ of the host's `patch' averaged
over the sphere $R$.

\subsection{Mass distribution function}
\label{sec:mass-df}

In interpretation of \citet{Bond--1991ApJ...379..440B} for the PS theory, the
excursion set formalism relies on two key ideas. First, only those virialized
haloes are counted for mass function, which are on a top level of a
hierarchical tree, i.e. not sub-haloes of any other halo. Second, the
conditions for fluctuations to form a virialized object may be implied to
linear stage of their evolution, and with a good apporximation they are
conditions for the filtered overdensity only. These ideas were successfully
realized in terms of a random process for the overdensity $\phi(R)$ filtered
over the scale $R$ acting as a parameter of the random process. The process
starts at the value $0$ and the parameter $R = \infty$ moving toward
$R = 0$. The mass function can be expressed then via the distribution function
for the parameter values when the process first crossed a certain threshold
$\delta_\mathrm{c}$. In
the absence of constraints the solution for this problem is the PS distribution
function (to the correcting factor `two'):
\begin{equation}
  \label{eq:variance-pdf-ps}
  f_S = \frac{\delta_\mathrm{c}}{\sqrt{2\pi S^3\,}}\,
  \exp\left( -\frac{\delta_\mathrm{c}^2}{2 S} \right)  \;,
\end{equation}
where the variance $S$ used as an equivalent measure of the filtering scale.
\citet{Bond--1991ApJ...379..440B} noted that this random process is not
markovian in general, but only if k-sharp filter used. In their paper the
authors proposed a procedure to calculate corrections to the PS distribution
function for a general filter (see also
\citet{Maggiore--2010ApJ...711..907M}). However, in this Paper we will neglect
corrections of such kind for simplicity.

In case of a general gaussian random process $\phi(S)$ the distribution
function can be written in a form \citep{Maggiore--2010ApJ...711..907M}
\begin{equation}
  f_S = - \DPfrac{}{S} \int \mathcal{D}[\phi]\,
  \operatorname{exp}\left( -\frac{1}{2}\,\phi^\mathrm{T} A^{-1} \phi \right)
  \;.
\end{equation}
This integral is computed over all possible trajectories of the field, not
exceedeing the threshold $\delta_\mathrm{c}$. The look of integration measure
$\mathcal{D}[\phi]$ is not important for our study, except it must be
positive. The covariance matrix in our case is just
\begin{align}
  A(r, S', S'')
  &= \langle \phi(r, S')\,\phi(r, S'') \rangle  \\
  &= 4\pi \int_0^\infty \mathrm{d}k\,k^2\,
  \tilde W(k, S')\,\tilde W(k, S'')\,P(k)
  - Q_{\alpha\beta}^{-1}\,\xi_\alpha(r, S')\,\xi_\beta(r, S'')  \;.
\end{align}
Assume firstly the case when $r = 0$ and
$\tilde H_\alpha(k) = \tilde W(k, S_\alpha)$. Denoting
$\xi(S', S'')
\equiv 4\pi \int_0^\infty \mathrm{d}k\,k^2\,
\tilde W(k, S')\,\tilde W(k, S'')\,P(k)$ we have
$Q_{\alpha\beta} = \xi(S_\alpha, S_\beta)$ and
\begin{equation}
  A(0, S', S'') = \xi(S', S'')
  - Q_{\alpha\beta}^{-1}\,\xi(S', S_\alpha)\,\xi(S'', S_\beta)  \;.
\end{equation}
It is obvious that $A(0, S', S_\alpha) \equiv A(0, S_\beta, S'') \equiv 0$,
i.e. the covariance matrix has zeroth rows and columns corresponding to each
constraining scale. Hence, the inverse of the covariance matrix is singular at
the parameters' pairs $(S_\alpha, S_\beta)$. For this reason the contribution
of the field to the intergral at this points is zero. Due to positiveness of
the integrand function, the last is also true for a certain vicinities of these
pairs. Thus, the distribution function $f_S$ should turn to zero when $S =
S_\alpha$. In a more general case, if $r > 0$, the covariance matrix does not
vanish but suppressed at the corresponding rows and columns. This also leads to
a suppression of the distribution function at the constraining scales. Indeed,
as the mean overdensity has been fixed at a scale $S_\alpha$, its variance
vanishes, so at this scale the structures does not form.

The cross-correlator function $\xi(S', S'')$ behaves roughly like
$\min(S', S'')$ (this is exactly true for the k-sharp filter). Hence, at large
spatial scales (small variance) the covariance matrix reduces to the
non-constrained one (and also the distribution function does). On the other
hand, at small spatial scales it tends to the non-constrained matix, minus some
constant amount of order $Q_{\alpha\beta}^{-1} S_\alpha S_\beta$. These
limiting cases and all the above suggest us the form of the distribution
function preserving the behaviours just investigated. It is the
Eq. (\ref{eq:variance-pdf-ps}), after substituting the constrained variance
$\Sigma$ instead of the non-constrained $S$. Taking this as an approximation
write out the final definition of the mass PDF which is adopted in subsequent
calculations:
\begin{equation}
  \label{eq:mass-pdf}
  f_m = \left| \DPfrac{S}{m} \right|
  \frac{\delta_\mathrm{c}}{\sqrt{2\pi \Sigma^3\,}}\,
  \exp\left( -\frac{\delta_\mathrm{c}^2}{2 \Sigma} \right)  \;.
\end{equation}
Here used the relation between mass and the filtering scale,
$m = (4\pi/3)\,\Omega_\mathrm{m,0} \rho_\mathrm{cr,0} R^3$. This mass PDF
should be renormalised to unity integral.

Let us summarize the properties of this function. First of all, it depends on
spatial position inside the host since the growth factor and the variance
depend too. These dependencies are defined by the constraints (or shape of the
host profile) as well as their values. Presence of the host itself decreases
the variance of the fluctuations' field.

Connection of the background mass distribution to clustering of haloes was
considered much earlier by \citet{Bond--1991ApJ...379..440B} and
\citet{Bower--1991MNRAS.248..332B}. Their biased mass PDF can be formally
defined as a result of substitutions
$\delta_\mathrm{c} \mapsto \delta_\mathrm{c} - \delta_\mathrm{H}$ and
$S \mapsto S - S_\mathrm{H}$ into the PS distribution
(\ref{eq:variance-pdf-ps}), where $S_\mathrm{H}$ is the non-constrained
variance on the scale of the host, and $\delta_\mathrm{H}$ is the linear
overdensity of the host. This result appears in the presented theory as a
special case, when the top-hat kernel is used both for single constraint
and for filtering, and when the PDF considered at the centre of the host. In
this case the variance reduces to the simple difference
$\Sigma = S - S_\mathrm{H}$, and the threshold overdensity can be written as
the difference noted above when it's expressed relatively to the global
density instead of the host's. According to previous authors this special case
will be named the extended PS (ePS) model.

\begin{figure*}
  \begin{center}
    \includegraphics[width=\textwidth]{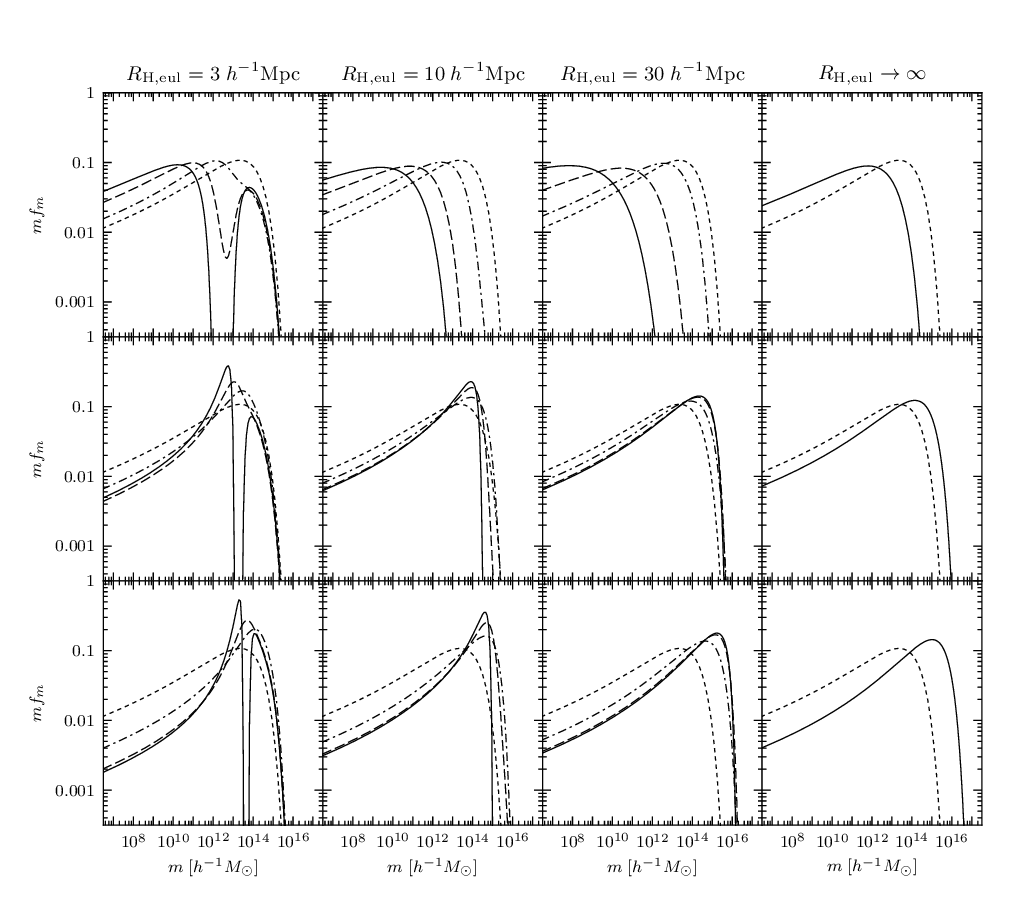}
  \end{center}
  \caption{Mass PDF (multiplied by $m$) at $z = 0$ for a single top-hat
    constraint having different scales and volume-averaged overdensities. Each
    row correspond to a certain overdensity $\overline\Delta$ (from top to
    bottom): $-0.65$, $1$, $4$. Columns from 1 to 3 correspond to a certain
    constraining scale $R_\mathrm{H}$ (from left to right): $3$, $10$, $30$
    $h^{-1}$Mpc. On each plot the PDFs is shown for $r = 0$ (solid line),
    $r = R_\mathrm{H}/2$ (long-dashed line), and $r = R_\mathrm{H}$ (dot-dashed
    line). Plots in column 4 correspond to `uniform' case (see text). The
    short-dashed line stands for non-constrained mass PDF of Press\&Schechter.}
  \label{fig:mass-pdf}
\end{figure*}

Resulting mass PDFs for the single top-hat constraint are shown at
Fig. \ref{fig:mass-pdf}. These runs differ by the volume-averaged
overdensity and the eulerian radius of the host (according to
Eq. (\ref{eq:lagrangian-to-eulerian})). At the first three columns the eulerian
radii of the hosts are finite, hence, the effect of modes correlation is on
hand appearing as the radial dependence of the shape of the PDF.
The most meaningful constrasts between the constrained PDFs and PS one
(short-dashed line) is seen for PDFs measured at the centre of the host (solid
line), while the outer area of the host has a lesser impact on the modes
statistics. Dips and gaps correspond to the mass scale of the entire host.
While the gaps are likely artefacts of the spherical top-hat constraint, the
certain suppression of the PDF might be a common feature marking the mass scale
of the host (e.g. this is the feature of the derivation constraints and momenta
constraints also, see Fig. \ref{fig:variance} and Subsection \ref{sec:kernels}).
Qualitatively the same result was obtained by
\citet{Sheth--1999MNRAS.308..119S} where PDF is cut on the host mass scale.

The rightmost column is for a formally `uniform' limit
$R_\mathrm{H} \to \infty$, when the statistics of modes reduces to the
non-constrained case, so the population of haloes is affected only through the
growth factor. This is also the low mass limit for the PDF.

From the observational point of view the number density of haloes of certain
mass is more preferable value than the mass PDF. The number of haloes per unit
comoving volume (i.e. the lagrangian one) and unit mass is
\citep{Sheth--1999MNRAS.304.767S}
\begin{equation}
  n(z, r, m)
  = \Omega_\mathrm{m,0} \rho_\mathrm{cr,0}\,\frac{f_m(z, r, m)}{m}  \;.
\end{equation}
Binding to observational volume, i.e. to the eulerian one, should be done with
jacobian of coordinate transformation. In the case of the spherical host the
transformation is (\ref{eq:lagrangian-to-eulerian}), so
$n_\mathrm{eul} = (1 + \Delta)\,n$.

\begin{figure*}
  \begin{center}
    \includegraphics{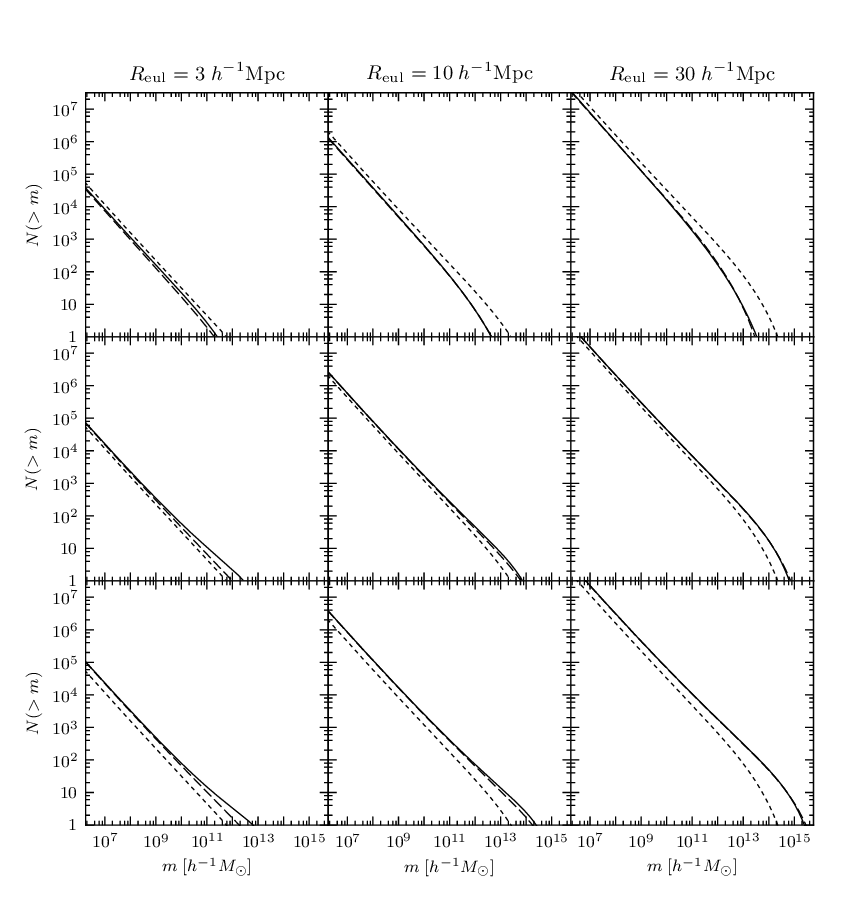}
  \end{center}
  \caption{Cumulative mass function. Columns are for three values of the
    sphere of interest, and rows are for three values of the mean overdensity
    of the host (the same as Fig. \ref{fig:mass-pdf}). Solid lines show the
    mass CDF in cases when radius of the sphere of interest coincides with the
    constraining radius. Long-dashed lines show the models where the
    constraining radius is infinite, $R_\mathrm{H} \to \infty$. Short-dashed
    lines show results of the non-constrained runs.}
  \label{fig:mass-cdf}
\end{figure*}

In the cumulative mass function all the strong features of the PDF, like dips,
are smoothed when integrating over the volume. The cumulative number of haloes
inside a sphere of a given lagrangian radius is
\begin{equation}
  \label{eq:mass-cdf}
  N(z, <R_\mathrm{lag}, >m) =
  4\pi \int_0^{R_\mathrm{lag}} \mathrm{d}r\,r^2
  \int_m^{m_\mathrm{lag}} \mathrm{d}{m'}\,n(z, r, m')  \;,
\end{equation}
where
$m_\mathrm{lag}
= (4\pi/3)\,\Omega_\mathrm{m,0} \rho_\mathrm{cr,0} R_\mathrm{lag}^3$ is the
mass enclosed in the sphere of interest with lagrangian radius
$R_\mathrm{lag}$. On Fig. \ref{fig:mass-cdf} depicted the mass CDF for
different constraining scales and overdensities.
Three classes of models are
considered: first where the constraining radius and radius of the sphere of
interest are coincide (solid lines), second one when the `uniform' background
is used, i.e. constraining radius is infinite (long-dashed lines), and third
one correspond to the non-constrained runs. The differences between first two
are negligible, except the small sphere of interest case. Deviations from the
non-constrained run is significant in all runs. It's easy to show that on a
small scales
($m_\mathrm{lag} \ll M_\mathrm{H} \equiv
(4\pi/3)\,\Omega_\mathrm{m,0} \rho_\mathrm{cr,0} R_\mathrm{H}^3$) such a
deviation depends on the mean host overdensity only. This behaviour of the mass
CDF was well known before as a `bias' \citep{Mo--1996MNRAS.282..347M}.

It is important to address the question of scatter in the distribution
functions specified above. The scatter can be of two reasons: a deviation of
the fluctuations field around the mean profile, and a shot noise in halo
counting.

\citet{Munoz--2008MNRAS.391.1341M} introduced a method for estimation of the
supercluster overdensity in ePS model. Their method based on an assumption that
given the host overdensity the number of sub-haloes are distributed accordingly
to Poisson law. Connection to the host overdensity was provided with the
equation equivalent to (\ref{eq:mass-cdf}) but for the mass PDF $f_m$ by
\citet{Barkana--2004ApJ...609..474B}. Their model allowed to estimate the
sub-haloes number as well as the overdensity scatter inside the host. Let's
evaluate the last for a more general case.
Define a distance in a space of profiles using the
inversion of the correlation `matrix' (\ref{eq:correlation-function}) as a
metric. The square of distance between a trial profile $\delta$ and the mean is
then
\begin{equation}
  \mu^2
  = (4\pi)^2 \int_0^\infty \mathrm{d}k\,k^2
  \int_0^\infty \mathrm{d}k'\,k'^2\,K^{-1}(k, k')
  \left( \tilde{\delta}(k) - \tilde{\Delta}(k) \right)
  \left( \tilde{\delta}(k') - \tilde{\Delta}(k') \right)  \;,
\end{equation}
Can be seen that the `matrix' $K(k, k')$ has very strong diagonal so
approximately we can write
\begin{equation}
  K^{-1}(k, k')
  \approx \frac{\delta_\mathrm{1D}(k - k')}{4\pi k k'}\,\frac{1}{P(k)}  \;.
\end{equation}
If we determine the trial profile in terms of constraints with kernels $h_i$,
constraints' correlation matrix $q_{ij}$, and constraints' values  $c_i$, then
the squared distance can be written as
\begin{multline}
  \mu^2
  = q_{ij}^{-1} c_i c_j + Q_{\alpha\beta}^{-1} C_\alpha C_\beta  \\
  - 2 q_{ij}^{-1} c_i Q_{\alpha\beta}^{-1} C_\alpha\,
  4\pi \int_0^\infty \mathrm{d}k\,k^2
  \tilde h_j(k)\,\tilde H_\beta(k)\,P(k)  \;.
\end{multline}
Given the confidence level $\mu$ it is possible to restrict the values $c_i$
so the acceptable shapes.
As a particular case, if let the profile's deviation to be strictly
proportional to the mean profile, the proportionality factor appropriate to the
confidence level $\mu$ is then
$\mu / (Q_{\alpha\beta}^{-1} C_\alpha C_\beta)$. In a purely ePS model (which
is equivalent to the case with single top-hat constraint, see above) it's just
$\mu \sqrt{S(R_\mathrm{H})} / \overline{\Delta}(R_\mathrm{H})$.
Accordingly
to Fig. \ref{fig:variance}, this kind of scatter should not matter much in
hosts with enclosed mass $\gtrsim 10^{15}$~$h^{-1} M_\odot$ or equivalent
lagrangian scale $\gtrsim 10$~$h^{-1}$Mpc.

Another source of scatter arises from the random nature of the excursion
process itself. A clustering process was investigated by
\citet{Sheth--1996MNRAS.281.1277S} then enhanced by
\citet{Sheth--1997MNRAS.289...66S} and
\citet{Sheth--1999MNRAS.304.767S}. The approach cited lead to results
equivalent to the ePS model \citet{Sheth--1996MNRAS.281.1277S}.
Let interpret the value of the cumulative mass function $N(>m)$,
Eq. (\ref{eq:mass-cdf}), as an ensemble mean of a random function
$\mathcal{N}(>m)$, which may be represented as a sum over bins of the mass
greater than the given, i.e. $\mathcal{N}(>m) = \sum_i \mathcal{N}_i$.
Variance of last is then
\begin{equation}
  \label{eq:mass-cdf-variance-general}
  \left\langle (\mathcal{N}(>m) - N(>m))^2 \right\rangle =
  \sum_{i,j} \langle \mathcal{N}_i \mathcal{N}_j \rangle -
  \sum_{i,j} \langle \mathcal{N}_i \rangle \langle \mathcal{N}_j \rangle  \;.
\end{equation}
Assuming ePS conditions (there are single top-hat constraint defining a host of
a mass $M_\mathrm{H}$) and standing on the Poisson initial distribution
approximation \citep{Sheth--1996MNRAS.281.1277S}, the cross-correlator
$\langle \mathcal{N}_i \mathcal{N}_j \rangle$ for $i \neq j$ can be expressed
in a form similar to eq. (A46) of \citet{Sheth--1999MNRAS.304.767S}. For
$i = j$ it can be written using their eq. (A49) for $\alpha = 2$. As shown by
\citet{Sheth--1999MNRAS.304.767S}, in a small mass limit, $m \ll M_\mathrm{H}$,
the cross-correlators decay like
$\langle \mathcal{N}_i \mathcal{N}_j \rangle \approx
\langle \mathcal{N}_i \rangle \langle \mathcal{N}_j \rangle$ for $i \neq j$,
and also $\langle \mathcal{N}_i^2 \rangle \approx
\langle \mathcal{N}_i \rangle^2 + \langle \mathcal{N}_i \rangle$. In the
opposite limit, $m \sim M_\mathrm{H}$, the cross-correlators vanish as well as
$\langle \mathcal{N}_i \rangle$. As a result, the poissonian estimation for
variance (\ref{eq:mass-cdf-variance-general}) can be quite reliable, i.e.
\begin{equation}
  \label{eq:mass-cdf-variance}
  \left\langle (\mathcal{N}(>m) - N(>m))^2 \right\rangle \lesssim N(>m)  \;.
\end{equation}

\section{Recovering profile parameters of a host}

As it turned out, the mass CDF can significantly depend on the host
parameters. Consider the possibility to solve the inverse problem, i.e. to
recover the host parameters through the measured cumulative mass function of
haloes. The target setting for this inverse problem may be as follows:
\begin{enumerate}
  \item[--] choose a non-virialized structure having spherical shape, or a
    spherical subvolume;
  \item[--] choose a set of basis profiles reflecting values we'd like to
    measure (see Subsec. \ref{sec:kernels});
  \item[--] measure the mass function of virialized substructures inside it,
    and/or values related to the constraining values;
  \item[--] adjust theoretical parameters of the host (lagrangian constraining
    scales and/or constraining values) to fit the theoretical data to the
    measured data.
\end{enumerate}
Evidently, this inverse problem may be targeted not only to recognize the mean
overdensity of the host but also shape and profile details, using a proper set
of the host parameters.

To test the method let's focus on a simple case, when we interested in only the
volume-averaged overdensity of the host. This problem was already investigated
before by \citet{Munoz--2008MNRAS.391.1341M} and
\citet{Sheth--2011MNRAS.417.2938S} as an application to the ePS model to
Shapley's supercluster and Sloan Great Wall. Consider the eulerian scale and
the mass function at a given redshift are the input parameters. The output will
be the host overdensity linearly evolved to the corresponding
redshift. Theoretical part herewith is reduced to the case of the single
top-hat constraint.

\begin{figure*}
  \begin{center}
    \includegraphics{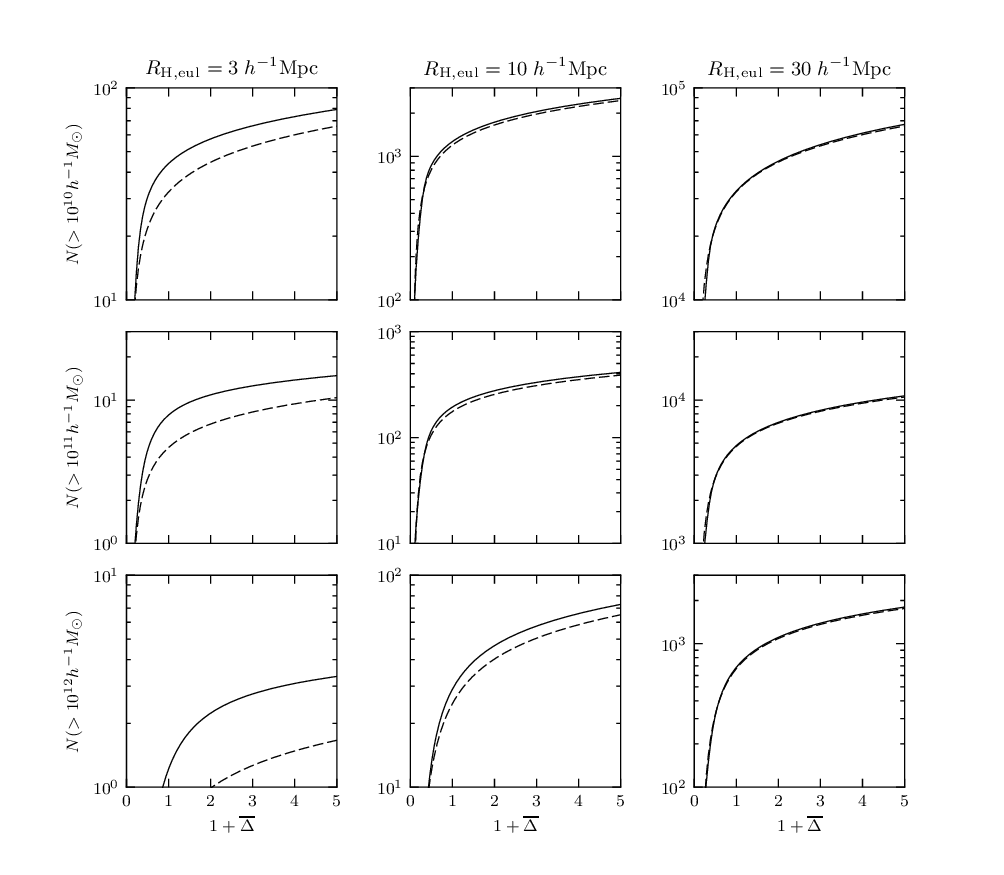}
  \end{center}
  \caption{Number of haloes having mass greater than given, as a function of
    averaged overdensity. The single top-hat constraint is used. Every column
    correspond to a certain eulerian radius where the mass CDF is computed
    (the sphere of interest). Solid lines show the case when radius of the
    sphere of interest coincides with the constraining radius. Long-dashed
    lines shows the `uniform' models (see text).}
  \label{fig:number-df-overdensity}
\end{figure*}

On Fig. \ref{fig:number-df-overdensity} is given the number of haloes depending
on lower halo mass in the sample, as well as the host overdensity and its
scale. Like on the Fig. \ref{fig:mass-cdf}, results of the runs for the finite
and infinite-sized hosts are indistinguishable in practice, except for the case
$R_\mathrm{H,eul} = 3$~$h^{-1}$Mpc where the boundary effects play crucial role.
As seen, the mapping of the overdensity to the haloes number is
not degenerated, though the low density limit is more reliable in sense of
the inverse problem.

\begin{figure}
  \begin{center}
    \includegraphics{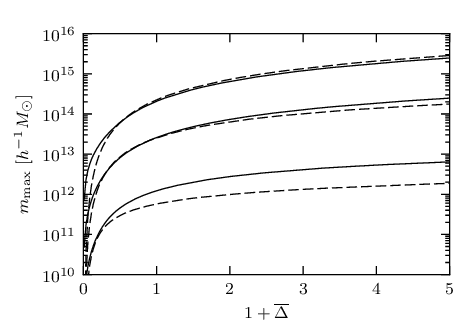}
  \end{center}
  \caption{Maximum mass of a halo containing in a host of given eulerian
    radius, from top to bottom: $30$~$h^{-1}$Mpc, $10$~$h^{-1}$Mpc,
    $3$~$h^{-1}$Mpc. The styles of lines denote the same as on
    Fig. \ref{fig:number-df-overdensity}.}
  \label{fig:mass-df-overdensity}
\end{figure}

To successfully apply the method described above, the full lower-bounded by
mass sample of haloes is necessary. Another way is to adopt a maximum mass halo
to bind the overdensity. As seen on Fig. \ref{fig:mass-df-overdensity} it is
quite sensitive indicator, except for high density case where the represented
dependency effectively degenerated.

Both methods for solving the inverse problem are more sensitive with respect to
voids rather than superclusters. On the other hand, observational difficulties
can somewhat complicate the harvesting of statistics for haloes
population. Such difficulties are large spatial extent together with small
surface brightness of void galaxies. In spite of this, the methods provided can
be helpful for investigation of the problem of voids and superclusters.

\section{Discussion and conclusion}

In this Paper was presented the method for describing the evolution of a
virialized halo population in superclusters and cosmological voids. The method
is based on the model of \citet{Press--1974ApJ...187..425P} in
interpretation of \citet{Bond--1991ApJ...379..440B} well known as the excursion
set formalism. The difference between last and presented model is that the host
is described in terms of the statistical constraints implied to the initial
overdensity fluctuation field. The constraining procedure used is the method of
\citet{Hoffman--1991ApJ...380L...5H,Hoffman--1992ApJ...384..448H} reformulated
for spherical harmonics. The constraints have form of the linear functionals,
and its kernels determine spatial scale and shape of the host perturbation
which evolves to the supercluster or void. The model accounted for explicit
positional dependency of statistical characteristics of the overdensity
field. As a result of imposing the constraints, the characteristics of the
statistical field, as well as the  halo mass function, clearly become spatially
related to the parameters of the background structure. The statistical
constraints enable us to specify parameters of the background structure such as
the mean density, moments of inertia, gradients, etc.

In the particular cases considered above, the background structure was assumed
to be spherically symmetric. However, it is not difficult to generalize our
formalism to a structure of arbitrary shape, e.g., a cosmological wall or a
filament.

As an application, we have considered the recovery of the mean density of the
background structure using both the observed integrated mass function and the
mass of the most massive halo. Both formulations yield results that are more
sensitive to cosmological voids than superclusters. On the other hand, the fact
that it is difficult to observe galaxies within voids makes the acquisition of
reliable halopopulation statistics more complex, since voids are large and
sparsely filled, and galaxies within the voids have low surface
brightnesses. Nevertheless, this method may be helpful in the study of
superclusters and voids.

\acknowledgements

This work has been funded by grant of the President of Russian Federation for
supporting of scientific Schools NSh-3602.2012.2 and also by grant
N14.А18.21.1179.

\end{document}